\documentclass[aps,prb,twocolumn]{revtex4}
\usepackage{bm,hyperref}
\usepackage{amsmath}
\usepackage{graphicx}
\usepackage{soul}
\usepackage{amsfonts}
\usepackage{amssymb}
\usepackage{dcolumn}
\usepackage{amsmath}
\usepackage{esint}

\def\-#1{{\bf #1}}

\begin{document}
\title{Optical alignment of oval graphene flakes}
\author{E. Mobini{$^{*,1}$}, A. Rahimzadegan{$^{2}$}, R. Alaee{$^{2,3}$}, and C. Rockstuhl{$^{2,4}$}}
\address{
{$^{1}$}Abbe Center of Photonics, Friedrich-Schiller University, Jena, Germany\\
{$^{2}$}Institute of Theoretical Solid State Physics, Karlsruhe Institute of Technology, Karlsruhe, Germany\\
{$^{3}$}Max Planck Institute for the Science of Light, Erlangen, Germany\\
{$^{4}$}Institute of Nanotechnology, Karlsruhe Institute of Technology, Karlsruhe, Germany\\
$^*$ehsan.mobini@uni-jena.de
}

\begin{abstract}
Patterned graphene, as an atomically thin layer, supports localized surface plasmon-polaritons (LSPPs) at mid-infrared or far-infrared frequencies. This provides a pronounced optical force/torque in addition to large optical cross sections and will make it an ideal candidate for optical manipulation. Here, we study the optical force and torque exerted by a linearly polarized plane wave on circular and oval graphene flakes. Whereas the torque vanishes for circular flakes, the finite torque allows rotating and orienting oval flakes relative to the electric field polarization. Depending on the wavelength, the alignment is either perpendicular or parallel. In our contribution, we rely on full-wave numerical simulation but also on an analytical model that treats the graphene flakes in dipole approximation. The presented results reveal a good level of control on the spatial alignment of graphene flakes subjected to far-infrared illumination.
\end{abstract}

\maketitle
Graphene, a two dimensional crystal of carbon atoms arranged in a honeycomb pattern,
exhibits intriguing photonic and electronic properties~\cite{novoselov2005two,novoselov2004electric}. Dynamical tuning of the conductivity via gate voltage or chemical doping~\cite{vakil2011transformation},
a tunable bandgap via electrical gating~\cite{wang2008gate,zhang2009direct},
a higher level of light confinement compared to plasmonic materials~\cite{jablan2009plasmonics,mikhailov2007new,hwang2007dielectric},
and a high ratio of extinction cross section to the geometrical cross
section~\cite{thongrattanasiri2012complete} are a few to mention. Among the aforementioned
properties, it is most notably the possibility to enhance the light-matter interaction that provides enough motivation to consider graphene in various photonic applications. To observe a resonant light-matter interaction one requires to nano and/or micro pattern graphene with suitable shapes, e.g. ribbons~\cite{fei2015edge,piper2014total,Alaee:12} or disks~\cite{thongrattanasiri2012complete} such that it sustains localized surface plasmon polaritons. Applications emerging from the enhanced light-matter interaction would additionally benefit from the ability to optically manipulate the spatial position, arrangement, and orientation of the graphene flakes on demand comparable to conventional plasmonic particles~\cite{tong2009alignment,xu2016scattering,nome2009plasmonic,Juan2011}. Examples for such applications are optically reconfigurable materials~\cite{twombly2013optical},
trapping of micro/nano entities~\cite{zhang2016towards}, manipulating of dielectric particles~\cite{yang2016optical},
or optomechanical manipulation~\cite{mousavi2014strong}.

\begin{figure}
\begin{centering}
\includegraphics[scale=0.4]{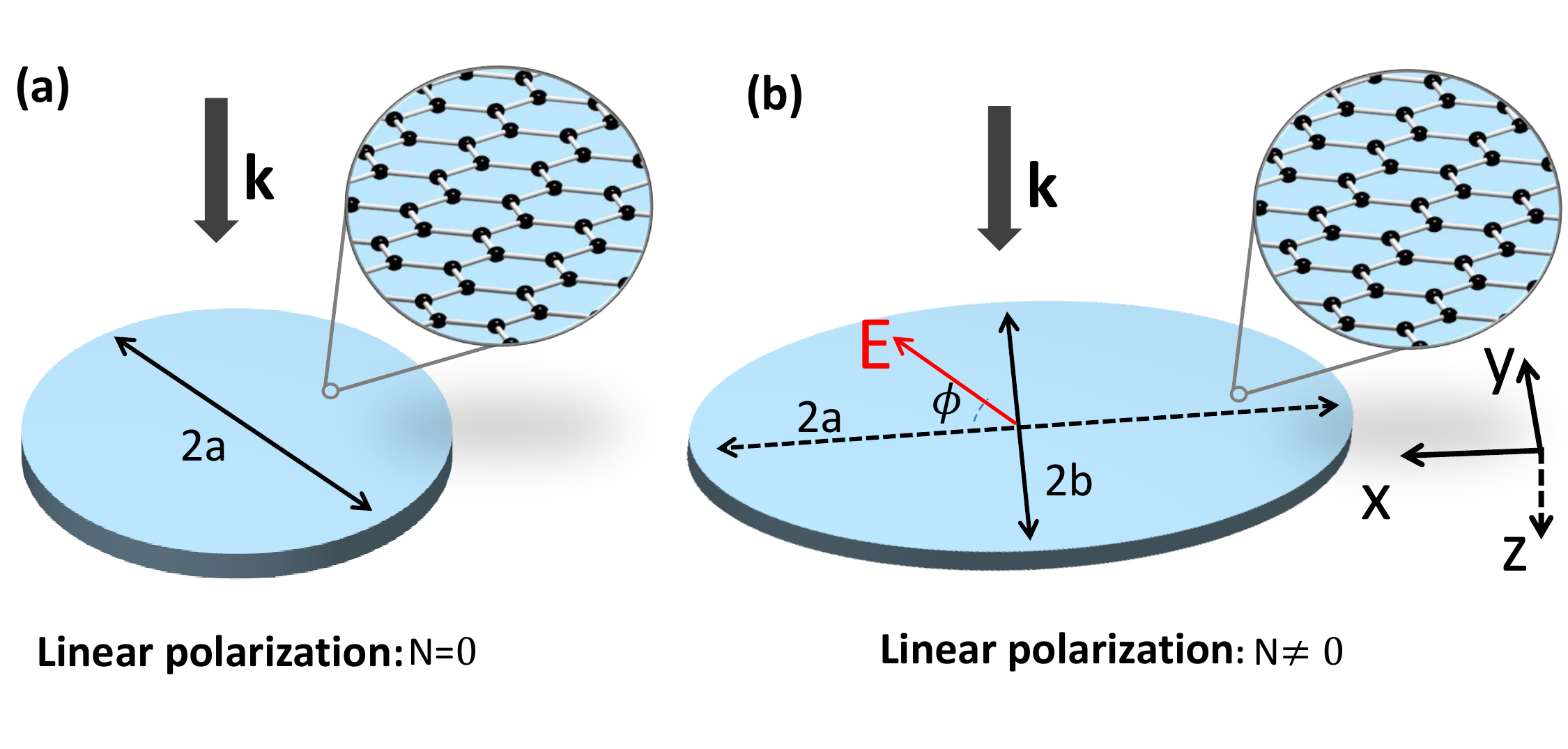}
\par\end{centering}

\protect\caption{Main idea of our work: A \textit{linearly} polarized plane wave can align and rotate an oval graphene flake. The figures on the left and right show the schematic of a circular graphene (in-plane isotropic) flake and an oval graphene (in-plane anisotropic) flake, respectively. Both are illuminated by a linearly polarized plane wave propagating in the $+z$ direction, perpendicular to the flake plane. The red arrow shows the electric field vector of the illuminating light on the flake plane.}\label{fig:1}
\end{figure}

Here, we study oval graphene flakes that, in contrast to circular graphene flakes, owing to their in-plane anisotropy, can be rotated by linearly polarized light. They can be aligned either parallel or perpendicular to the incident electric field vector, depending on the frequency of operation. The misalignment angle $\phi$ between the incident electric field and the major oval axis (here $+x$) can be tuned to control direction and magnitude of the exerted torque. This torque allows to align flakes upon request (Fig.~\ref{fig:1}).

In the following, we employ three approaches to calculate the polarizability, the optical cross section and the optical force and torque. Not each of these method is applied to each sub-aspect; but all together they provide a solid methodological framework to explore the properties of the pertinent system.

In the first approach, we study the full-wave dynamics in the entire setup. This requires to solve Maxwell's equations numerically. We use a finite element method (FEM) for this purpose~\cite{multiphysics2012}. The simulated fields are used to calculate the Maxwell's stress tensor (MST) from which eventually the {\it force and torque} as expressed in Eqs.~\ref{eq:a}-~\ref{eq:c} can be calculated. The results of this approach are exact. In the second approach (semi-analytic force and torque), we assume that the graphene flake possesses only an electric dipole response. This allows to use existing %CR: Instead of existing the word exciting was used. I felt it a bit of an exaggeration and replaced it with existing.
analytical expressions for the force and toque as expressed in Eqs.~\ref{eq:4}-~\ref{eq:6}. The approach is semi-analytic since a multipole expansion of the numerically simulated induced electric current density~\cite{jackson1999classical,fernandez2015exact} is used to extract the electric dipole moment of the graphene flake. In a third approach, in addition analytical expressions for the polarizability of the graphene flakes are used to compute the induced force and toque (quasi-static approximation). The agreement of the predictable force and torque with the different methods is assessed.  All calculations consider the graphene flakes to be in air. The electromagnetic quantities in phasor form have a time dependency of  $\exp(- \mathrm{i} \omega t)$. Quantities in time domain are denoted by an underline.

The time averaged mechanical force exerted on an arbitrary particle by an optical wave is calculated as~\cite{jackson1999classical,novotny2012principles}:

\begin{equation}
 \mathbf{F} =\left\langle \oint_S   \overset{=}{\underline{\mathbf{T}}}(\mathbf{r},t)\cdot\mathbf{n}\,\mathrm{d}S\right\rangle, \label{eq:a}
\end{equation}

\noindent where $S$ is any closed surface surrounding
the particle, $\mathbf{n}$ is a unit vector that points outward, and $\overset{=}{\underline{\mathbf{T}}}$ is the Maxwell's stress tensor.
The Maxwell's stress tensor is a tensor of second rank whose components can be calculated
as \cite{jackson1999classical,novotny2012principles}:
\begin{equation}
\underline{T}_{ij}=\varepsilon_{0}\left[\underline{E}_{i}\underline{E}_{j}+c^{2}\underline{B}_{i}\underline{B}_{j}-\frac{1}{2}\delta_{ij}\left(|\underline{\mathbf{E}}|^{2}+|\underline{\mathbf{B}}|^{2}\right)\right],\label{eq:b}
\end{equation}
\noindent in which $E$ and $B$ are the total (incident and scattered) electric and magnetic fields in the $i,j=x,y, z$ coordinates; and $\delta_{ij}$ is the Kronecker delta function.

The time averaged optical torque on an arbitrary particle by an optical wave can be calculated as:

\begin{equation}
\mathbf{N}=-\left\langle \oint_S \mathbf{n}\cdot\overset{=}{\underline{\mathbf{T}}}(\mathbf{r},t)\times\mathbf{r}\, \mathrm{d}S\right\rangle. \label{eq:c}
\end{equation}

This approach provides exact solutions but it complicates the physical discussion. To entail such discussion, we also apply a multipole expansion method, to expanded the induced current density in the graphene flakes into elementary multipole moments. The link between the incident field and the induced multipole moments is given by polarizability tensors.  For the oval graphene flake, that have a wavelength much longer than the size of the flakes, we can restrict our attention in good approximation to the electric dipole polarizability. The electric dipole in-plane polarizabilities of the flake can be expressed as:
\begin{equation}
\overset{=}{\alpha}= \alpha_{\Vert}\mathbf{e}_{x}\mathbf{e}_{x}+\alpha_{\bot}\mathbf{e}_{y}\mathbf{e}_{y},\label{eq:polar}
\end{equation}

with $\alpha_{\Vert}$ and $\alpha_{\bot}$ being the in-plane $x$ (parallel to the flake major axis) and $y$ (perpendicular to the flake major axis) polarizabilities, respectively.

For an electric dipolar particle (i.e. a particle with only a non-negligible electric dipole response)
illuminated with an arbitrary illumination, the induced optical force reads as \cite{chaumet2000time,nieto2015optical,chen2011optical,gusynin2006magneto,Rahimzadegan:PRB2016}:

\begin{equation}
\mathbf{F}_{p}=\frac{1}{2}\Re(\nabla\mathbf{E}^{*}\cdot\mathbf{p}),\label{eq:4}
\end{equation}

\noindent where $\mathbf{p}=\varepsilon_{0}\overset{=}{\alpha}\cdot\mathbf{E}$ denotes the induced
Cartesian electric dipole moment and $\overset{=}{\alpha}$ is the electric polarizability tensor
of the particle. If the particle is illuminated with a time harmonic linearly polarized plane wave propagating in the $+z$ direction

 \begin{equation}
 \mathbf{E}= E_{0}\left(\cos\phi~ \mathbf{e}_{x}+\sin\phi~\mathbf{e}_{y}\right)e^{ikz},\label{eq:EField}
 \end{equation}
where the polarization vector is oriented at an angle $\phi$
relative to the $+x$ axis (major axis of the oval), the optical force is calculated as:

\begin{equation}
\mathbf{F}_{p}=\frac{k^{3}}{2\pi}F^\mathrm{norm}\left[\Im(\alpha_{\Vert})\cos^{2}\phi+\Im(\alpha_{\bot})\sin^{2}\phi\right]\mathbf{e}_{z},\label{eq:5}
\end{equation}
\noindent with $k$ the wavenumber and $F^\mathrm{norm}=({I_{0}}/{c})(\lambda^{2}/{2\pi})$ the
normalization for the optical force. $I_{0}=\varepsilon_{0}c\left|E_{0}\right|^{2}/2$
is the intensity of the illumination. The physical importance of $F^\mathrm{norm}$ is that  $3F^\mathrm{norm}$ is the upper bound for the exerted optical force on an isotropic electric dipolar particle by a plane wave \cite{rahimzadegan2016fundamental}.

Similarly, the time averaged optical torque exerted on the flake by the same field is derived as \cite{nieto2015optical,chen2011optical,Rahimzadegan:PRB2016}:

\begin{align}
\mathbf{N} & =\frac{1}{2}\left\{ \Re\left(\mathbf{p}\times\mathbf{E}_\mathrm{inc}^{*}\right)-\frac{k^{3}}{6\pi}\left[\frac{1}{\varepsilon_{0}}\Im\left(\mathbf{p}^{*}\times\mathbf{p}\right)\right]\right\}. %\nonumber
\end{align}

\noindent By using the incident electric field, i.e. Eq.~\ref{eq:EField} and the induced dipole moment, i.e. Eq.~\ref{eq:polar}, we obtain:

\begin{align}
\mathbf{N} & =\frac{2k^{3}}{\pi}N^\mathrm{norm}\sin2\phi\left[\frac{1}{2}\Re\left(\alpha_{\Vert}-\alpha_{\bot}\right)-\frac{k^{3}}{6\pi}\Im\left(\alpha_{\bot}\alpha_{\Vert}^{*}\right)\right]\mathbf{e}_{z},\label{eq:6}
\end{align}
\noindent where $N^\mathrm{norm}=({I_{0}}/{\omega})\left({\lambda^{2}}/{8\pi}\right)$ is
the torque normalization.  $3N^\mathrm{norm}$ is the upper-bound for the exerted torque on an isotropic electric dipolar particle by a plane wave  \cite{rahimzadegan2016fundamental}.

\begin{figure}
\centering{}\includegraphics[scale=0.33]{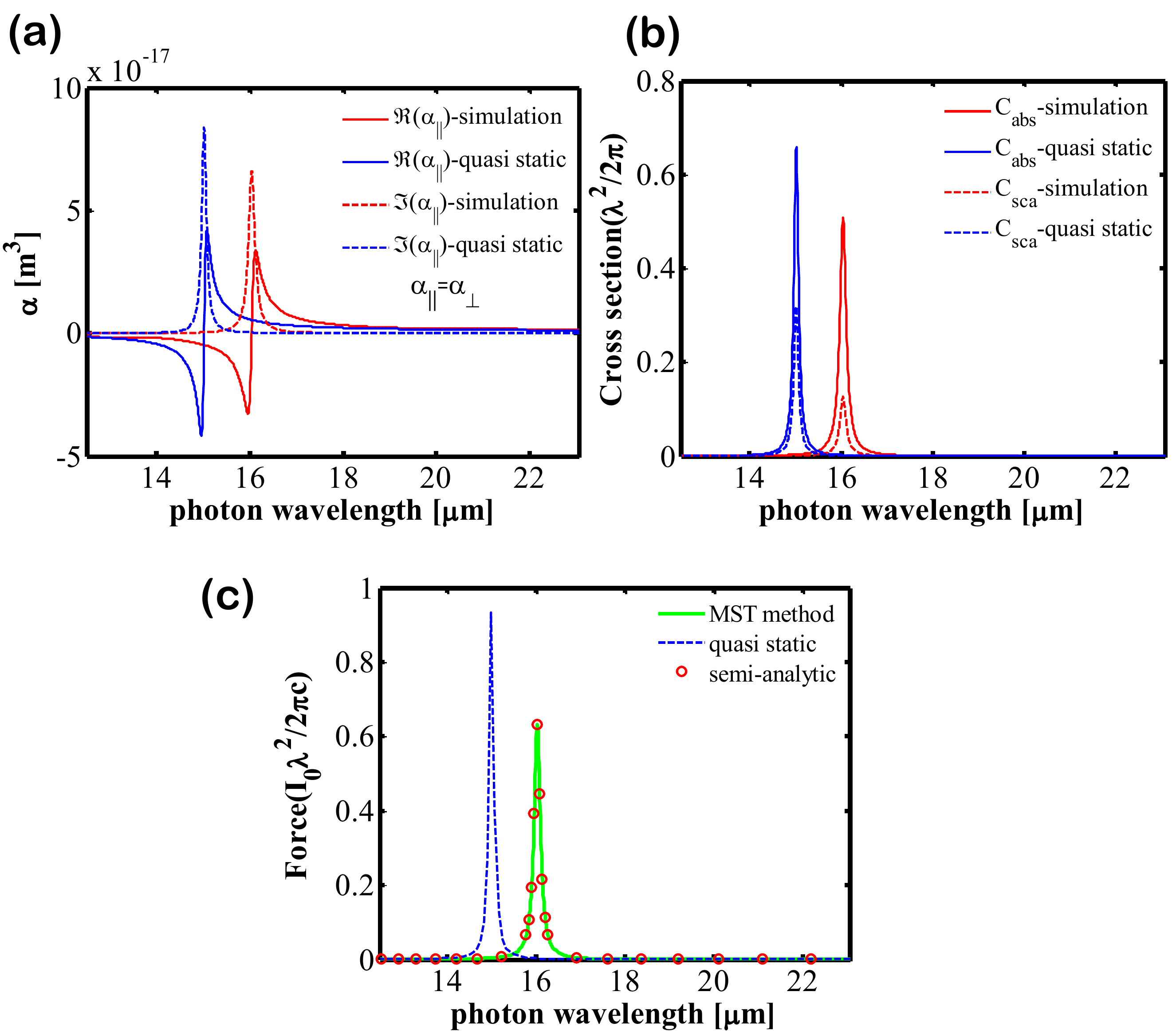}\protect\caption{\textit{Circular graphene flake} shown in Fig.~\ref{fig:1}a of diameter $D=1\,\mu m$. (a)
Real and imaginary parts of the electric dipole polarizability calculated with quasi-static method and FEM simulation. (b) The absorption and scattering cross sections of the flake upon plane wave illumination, using the polarizabilities calculated with the quasi static method and FEM. (c) The MST, semi-analytic and quasi static optical force exerted on the flake illuminated by a linearly polarized plane wave. }\label{fig:2}
\end{figure}

Now that the relations for the optical force and torque are known, in the next part, we will focus on calculating the polarizabilities. The graphene, here, is described through the surface conductivity ($\sigma_{GR}$) derived by a quantum mechanical approach known as the Kubo formula \cite{falkovsky2007optical,gusynin2006magneto,hanson2008dyadic}
and reads as:
\begin{eqnarray}
\sigma_{GR}(\omega)&=&\frac{\mathrm{i}e^{2}}{4\pi\hbar}\ln\left[\frac{2|\mu_{c}|-(\omega+i2\Gamma)\hbar}{2|\mu_{c}|+(\omega+i2\Gamma)\hbar}\right]  \\ &&+\frac{ie^{2}k_{B}T}{\pi\hbar^{2}(\omega+i2\Gamma)}\left[\frac{\mu_{c}}{k_{B}T}+2\ln\left(e^{-\frac{\mu_{c}}{k_{B}T}}+1\right)\right],\nonumber\label{eq:8}
\end{eqnarray}

\noindent where $e$, $\hbar$, $k_{B}$ are the universal constants for electron
charge, reduced Planck constant, and Boltzmann constant, respectively.
$T$ represents the temperature. $\mu_{c}$ and $\Gamma$ are physical
parameters of the graphene sheet to represent the chemical potential (or
Fermi energy $E_{F}$) and the intrinsic loss due to the charged particle
scattering, respectively. Graphene, here, is numerically modelled as
a thin layer of a dielectric with permittivity of $\varepsilon_{GR}=\varepsilon_{0}+i\frac{\sigma_{GR}}{\omega\Delta}$
where $\Delta=0.5\, nm$ is the thickness of dielectric \cite{vakil2011transformation}.
For energies well-below the Fermi energy (i.e. $\hbar\omega<E_{F}$
) and $k_{B}T \ll E_{F}$  the above formula can be reduced to the Drude conductivity:

\begin{equation}
\sigma_{GR}(\omega)=\frac{e^{2}}{\pi\hbar^{2}}\frac{iE_{F}}{\omega+i\tau^{-1}},\label{eq:9}
\end{equation}

\noindent where $\tau$ is the relaxation time($2\varGamma=\frac{\hbar}{\tau}$)
and is obtained based on the DC mobility dominated by impurities as
$\tau=\mu E_{F}/ev_{F}^{2}$ with mobility $\mu$ and Fermi velocity
$v_{F}\approx10^{6}\frac{m}{s}$  \cite{jablan2009plasmonics}.

In order to underpin a comparative study, let us start with the circular graphene flake (Fig.~\ref{fig:1}a) with a diameter $D=1\,\mu m$. In our case, this graphene disk is doped to a Fermi energy $E_{F}=1\, eV$ and
the DC mobility is $\mu=10000\, cm^{2}/Vs$. Besides, in pursuance
of a further qualitative understanding, the polarizabilities $\alpha_{\bot}$
and $\alpha_{\Vert}$ are found theoretically via the quasi-static
approximation and compared to those calculated by simulation. Considering
the dominant role of the dipolar plasmon mode, these polarizabilities
can be calculated in SI units as \cite{de2015plasmonics,garcia2013multiple,garcia2014graphene}:

\begin{equation}
\alpha_{\bot,\Vert}\approx4\pi D^{3}\frac{A}{L_{\bot,\Vert}-\frac{i(4\pi\varepsilon_{0})\omega D}{\sigma\left(\omega\right)}},\label{eq:7}
\end{equation}

\noindent where $D$ represents the characteristic length and is equal to the
diameter/square root of the area in the case of circular/oval flake.
The coefficients $A$ and $L_{\bot,\Vert}$ are constant and
only depend on the choosen geometry. The coefficients
$L_{\bot,\Vert}$ is calculated using the polarizability
formula for a perfectly conducting ellipsoid \cite{bohren2008absorption,landau2013electrodynamics}
where the dimension normal to the flake surface ($+z$) vanishes. For a circular graphene flake these
constants can be analytically found as $A={\pi}/{4}$, $L_{\bot}=L_{\Vert}={3\pi^{2}}/{2}$.
%CR: You wrote above that you wanted to start with the 'in-plane isotropic disk-shape graphene flake' (what a complicated term by the way) and now you make this gentle swift in the following to the elliposoidal particle. Reads strange, especially because no geometrical parameter was given at all (only much more later). These numbers are geometry independent? I don't think so. Maybe they don't depend on the absolut number of the length scales and only on their ration; but at least their ratio needs to be mentioned once. I changed the sentences but you have to put more efforts into the correct description.

The results of the calculated polarizability of the circular graphene flake are shown in Fig.~\ref{fig:2}a. We distinguish the situation where the polarizability is extracted from the induced current density as obtained from the FEM simulations and once analytically as just mentioned.

The results obtained with both methods are in qualitative agreement. We notice that a slight disagreement exists between the spectral positions of the resonances. This red-shift in the polarizabilities is a clear indication that we are slightly beyond the range of applicability for the quasi-static approximation. The agreement will be better the lower the ratio of $D/\lambda$. However, the dispersion is very well reproduced. Higher order multipole moments are not notably induced. This is a promising indication that the semi-analytical approach, that predicts observable quantities in dipole approximation once these moments have been extracted from full-wave simulations, is applicable.

Having the polarizabilities calculated, the scattering and absorption cross sections can be obtained by using:

\begin{equation}
C_{\mathrm{sca}} =  \frac{k^{4}}{6\pi}\left|\alpha\right|^{2},~~~~\\
C_{\mathrm{ext}}  =  k\mathrm{Im}\left(\alpha\right),~~~~\\
C_{\mathrm{abs}}  =  C_{\mathrm{ext}}-C_{\mathrm{sca}}.
\end{equation}

Using the simulated values of the polarizability, the optical cross sections are shown
in Fig.~\ref{fig:2}b. The circular graphene flake
shows a relatively strong peak around the wavelength of
$16\mu m$ at which the flake supports a plasmon resonance.
The optical force exerted on the
graphene flake by a linearly polarized plane wave propagating in the $+z$ direction is plotted in
Fig.~\ref{fig:2}c. The force is calculated with MST, semi-analytic and quasi static methods.
The orientation of the polarization is of no importance due to the symmetry of the flake.
For calculating the optical force in the semi-analytic method, the simulated polarizabilities are plugged into Eq.~\ref{eq:5}.  For the circular graphene flake, due to the symmetry, we have $\alpha_{\bot}=\alpha_{\Vert}$;
and, as expected, the induced torque vanishes (see Eq.~\ref{eq:6}).
As seen in the figure, both methods of MST and semi-analytic are in excellent agreement, indicating the dominant effect of the dipolar mode. The slight disagreement in resonance wavelength between the quasi-analytical theory and the semi-analytical methods also translates to the cross sections and the force.

\begin{figure}
\begin{centering}
\includegraphics[scale=0.32]{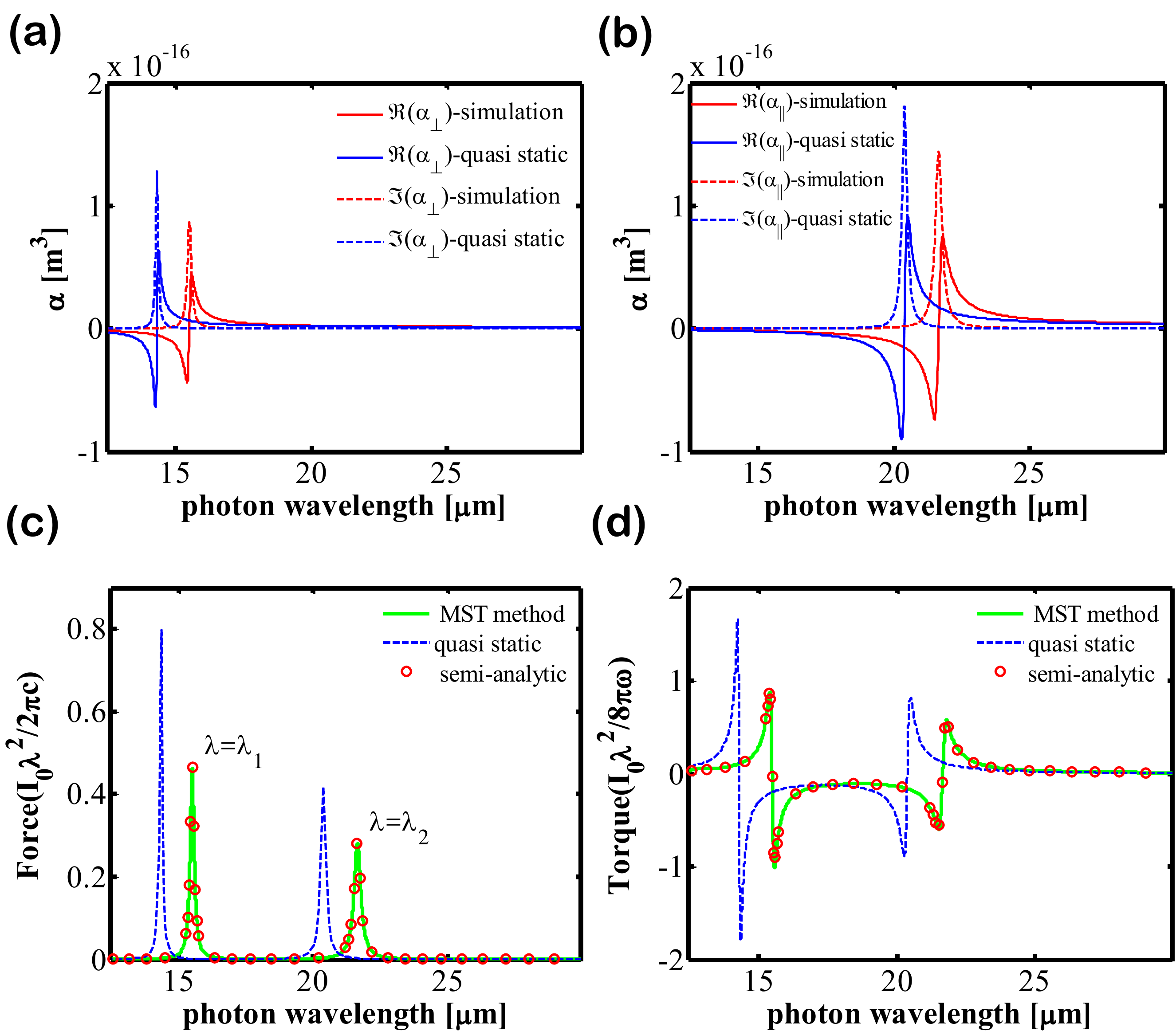}
\par\end{centering}
\protect\caption{\textit{Oval graphene flake } shown in Fig.~\ref{fig:1}b. The major and minor semi-axises have a length of $a=0.8\,\mu m$ and $b=0.5\,\mu m$, respectively. (a)
Real and imaginary parts of the electric dipole polarizability calculated with quasi-static approximation and FEM simulation parallel and (b) perpendicular to the major axis. (c) The MST, semi-analytic and quasi static optical force and (d) torque exerted on the flake by a linearly polarized plane wave illumination with a misalignment angle $\phi=\pi/4$. \label{fig:3}}
\end{figure}

To pursue the favoured aim of our analysis, we continue to investigate the same scenario
as above but now for the oval graphene flake
with a major and minor semi-axis of $a=0.8\,\mu m$ and $b=0.5\,\mu m$, respectively.
The misalignment angle $\phi$ that measures the orientation of the linearly polarized field relative to the major axis is set to $45^{o}$ (\ref{fig:1}b). The anisotropic geometry in this case causes the in-plane polarizabilities of the flake to take different values. For the oval graphene flake the constants in Eq.12 can be found by a suitable numerical integration as $A=1$, $L_{\bot}=18.2$,
$L_{\Vert}=9$. Interestingly, through these calculations, it can be
realized that these constants are not independent of each other and
the ratio ${L_{\Vert}}/{L_{\bot}}=\left({b}/{a}\right)^{\frac{3}{2}}$
holds true for an oval graphene flake.
Figures~\ref{fig:3}a and b show the in-plane polarizabilities parallel and perpendicular to the major axes. For each of the polarizabilities, there is a resonance occurring at wavelengths at which the other polarizability's magnitude is very
small. These resonance wavelengths can vary with the characteristic length of the object and the Fermi energy ($\lambda_\mathrm{res}\sim (D/E_{F})^{1/2}$
 \cite{garcia2014graphene}). Again, a slight disagreement between the simulated and the quasi-static polarizabilities can be seen but the dispersive features are well reproduced.

\begin{figure}
\centering
\includegraphics[scale=0.33]{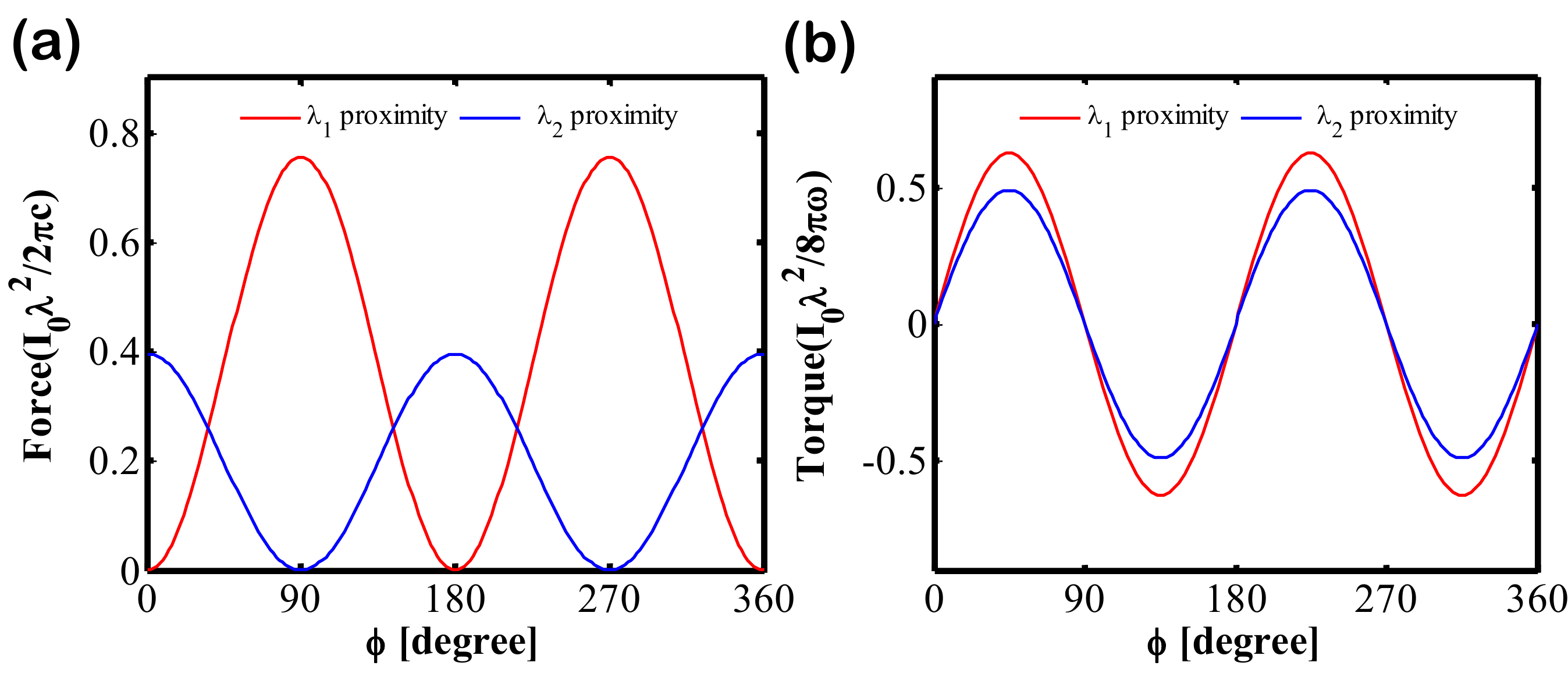}
\caption{\textit{Oval graphene flake}: (a) Calculated normalized optical force through the MST method at resonance wavelengths
$\lambda_{1}=15.44\,\mu m$ and $\lambda_{2}=21.7\,\mu m$ (b) calculated
normalized optical torque through the MST method around resonance wavelengths $\lambda_{1}$
and $\lambda_{2}$.\label{fig:4}}
\end{figure}

Figure~\ref{fig:3}c shows the calculated optical force on the flake with the three methods.
Two prominent resonances, calculated through the MST and semi-analytic methods, labeled with $\lambda_{1}$ and $\lambda_{2}$ correspond
to the resonances in each of the polarizabilities. The exerted optical torque is
illustrated in Fig.~\ref{fig:3}d. As is seen, the magnitude and sign varies
depending on the illumination wavelength. Figure~\ref{fig:3}d corroborates that a non-vanishing torque leads to a rotation with tendency of aligning the flake perpendicular, for $\lambda_{1}<\lambda<\lambda_{2}$ and
parallel, for $\lambda<\lambda_{1}$ and $\lambda>\lambda_{2}$, to the field polarization. The torque caused by the optical field attempts to minimize the potential energy of the object. Assuming the graphene flake is a dipolar particle, as noted above, its potential energy in terms of the induced dipole moment, $U=-\left\langle\mathbf{p}\cdot\mathbf{E}\right\rangle$, can be reduced to $-\left\langle\mathbf{p_{\bot}}\cdot\mathbf{E}\right\rangle$ and $-\left\langle\mathbf{p_{\parallel}}\cdot\mathbf{E}\right\rangle$ around $\lambda_{1}$ and $\lambda_{2}$ , respectively. Depending on the sign of these perpendicular and parallel polarizabilities, the torque directs the object towards the lower energy configuration i.e. parallel or perpendicular to the electric field. The optical force and torque
at $\lambda=\lambda_{1}$ and $\lambda=\lambda_{2}$ as a function of $\phi$ are illustrated in Fig.~\ref{fig:4}, showing that the exerted force and torque can be extended and modified by altering the incident field polarization.

In conclusion, we studied the exerted optical force and torque on an oval graphene flake and we demonstrated theoretically the possibility of its alignment and rotation by a linearly polarized plane wave illumination. Our findings show that for a specific oval graphene flake of particular size, by altering the Fermi energy, incident light wavelength and changing the misalignment angle ($\phi$), a good level of control on the direction and magnitude of rotation of the object is possible. This in turn allow us to achieve a tunable light manipulation of small size graphene flakes that can find applications in Micro-Opto-Electro-Mechanical Systems.

\end{document}